\documentclass{article}
\usepackage{spconf,amsmath,graphicx}
\usepackage{booktabs}
\usepackage{amssymb}
\usepackage{url}
\usepackage{float}
\usepackage{enumitem}

\title{Improving severity preservation of healthy-to-pathological voice conversion with global style tokens}
\name{Bence Mark Halpern$^{1,2}$, Wen-Chin Huang$^{1}$, Lester Phillip Violeta$^{1}$, R.J.J.H. van Son$^{2}$, Tomoki Toda$^{1}$}
\address{$^{1}$Nagoya University, Japan \\ $^{2}$Netherlands Cancer Institute, The Netherlands}

\copyrightnotice{979-8-3503-0689-7/23/\$31.00~\copyright2023 IEEE}
\begin{document}
\ninept
\maketitle
\begin{abstract}

In healthy-to-pathological voice conversion (H2P-VC), healthy speech is converted into pathological while preserving the identity. The paper improves on previous two-stage approach to H2P-VC where (1) speech is created first with the appropriate severity, (2) then the speaker identity of the voice is converted while preserving the severity of the voice. Specifically, we propose improvements to (2) by using phonetic posteriorgrams (PPG) and global style tokens (GST). Furthermore, we present a new dataset that contains parallel recordings of pathological and healthy speakers with the same identity which allows more precise evaluation. Listening tests by expert listeners show that the framework preserves severity of the source sample, while modelling target speaker's voice. We also show that (a) pathology impacts x-vectors but not all speaker information is lost, (b) choosing source speakers based on severity labels alone is insufficient.

\end{abstract}
\begin{keywords}
voice conversion, pathological speech, oral cancer speech, autoencoder
\end{keywords}
\section{Introduction}
\label{sec:intro}

Healthy-to-pathological voice conversion (H2P-VC) is a new voice conversion (VC) task, where healthy speech is converted so that it resembles characteristics of a speech pathology, while preserving the identity of the original speaker. H2P-VC could play a crucial role in several applications. For example, H2P-VC could be used to help patients mentally prepare for the changes in their future voice due to medical interventions \cite{huang2022towards, illa2021pathological, halpern2021objective}. By listening to the post-treatment voice, patients can better understand and adapt to the potential alterations, reducing anxiety and leading to a better quality of life \cite{Epstein1999}.  Additionally, H2P-VC could be also used as a data augmentation technique for improving automatic speech recognition (ASR) systems \cite{wang2023dutavc, harvill2021synthesis}. Moreover, H2P-VC holds potential for speaker anonymisation of clinically collected speech samples, making it valuable for protecting speaker identity in sensitive contexts, ensuring confidentiality while preserving severity that might be required for further analysis.

H2P-VC's main challenge is the lack of appropriate, parallel data, which leads to difficulties during evaluation, and means that only non-parallel VC techniques are applicable. The lack of parallel healthy and pathological data from the same speaker means that we do not have the ground truth (GT) for the task, which makes the evaluation of the speech complicated. Furthermore, most work on H2P-VC has been done using the UASpeech corpus \cite{kim2008dysarthric}, making it challenging to adequately model and evaluate severity features that might be apparent at the prosodic level.

Therefore, current approaches to H2P-VC resort to non-parallel VC techniques. These techniques either take an existing pathological speech sample \cite{illa2021pathological} or create a synthetic pathological speech sample \cite{huang2022towards}, which is then converted to a new speaker's identity. Recently, \cite{huang2022towards} proposed a two-stage model for the H2P-VC task.  The first stage focused on capturing time-variant aspects of the pathological speech using the Voice Transformer Network (VTN) \cite{huang20i_interspeech}. This led to the production of speech mimicking the severity of the pathological speech but, unfortunately, resulted in the loss of source speaker identity. During the second stage, \cite{huang2022towards} attempted to reconstruct the original identity of the healthy speaker using a VQVAE-based model \cite{van2017neural} from the \textit{crank} toolkit \cite{kobayashi2021crank}. Although the framework successfully altered the source speaker's characteristics, the result did not accurately resemble the target speaker's characteristics. The work hypothesised that these shortcomings were partially due to evaluation limitations, as \cite{huang2022towards} lacked access to parallel GT. An additional issue was that the second stage of the VC adversely affected the severity features of the speech.

\begin{figure}
\includegraphics[width=\columnwidth]{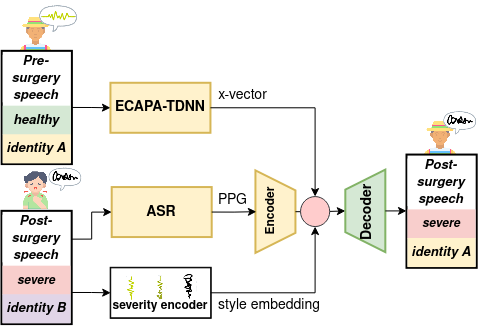}
\caption{Outline of our proposed approach for severity-preserving voice conversion.}
\label{fig:outline}
\end{figure}

To address these limitations, our study's primary focus is enhancing the second stage of the method presented in \cite{huang2022towards}. We propose a new VC model leveraging phonetic posteriorgram (PPG) and global style tokens (GST) shown in Figure~\ref{fig:outline}. Many PPG-based VC systems were among the best in the VCC2020 challenge \cite{yi20_vccbc}, however, they have a tendency to remove the severity features of the speech \cite{serrano2019parallel, zheng2022phonetic}, which is undesired in H2P-VC. Our approach aims to improve the preservation of the severity while effectively converting to the target speaker's voice. GSTs have been shown to be able to control prosody in text-to-speech (TTS) \cite{wang2018style}, and in VC \cite{liu2020transferring}. Using both GST and PPG, allows us to model the time-invariant (severity) and time-variant (intelligibility) aspects of the speech pathology separately, preserving the features on both levels.

In addition, we perform a more rigorous evaluation of severity, and speaker identity, using NKI-OC-VC, a newly collected oral cancer speech dataset, which contains parallel pathological and healthy recordings from the same speaker. Compared to the UASpeech corpus \cite{kim2008dysarthric}, the dataset contains utterance-level stimulus that is more appropriate to evaluate aspects of severity that might be only present in longer stretches of speech. Using this dataset, we aim to revisit an assumption of H2P-VC -- whether pathology impacts x-vectors. This assumption holds key importance. If a pathology indeed changes x-vectors notably, it directly impacts the effectiveness of our VC model. 

Furthermore, we also seek to reevaluate the current method of choosing source speakers for VC, which is based on matching severity ratings between the source and target speakers. We hypothesise that the same severity rating does not always mean the same type of voice changes. 

The primary contributions of our paper are the following: 
\begin{enumerate}[label=\textbf{(\arabic*)},noitemsep]
    \item We propose a new model for the second stage of voice conversion (VC) in \cite{huang2022towards}, which improves the preservation of speech severity while being able to convert identity.
    \item Using a unique, parallel oral cancer speech dataset, NKI-OC-VC, we present a comprehensive and precise evaluation of our VC system by speech-language pathologists (SLPs). 
    \item We reevaluate the assumption of voice pathology's impact on speaker's embedding and its implications on the accuracy of our VC model. 
    \item We ask speech-language pathologists (SLPs) to examine our current practice of matching severity ratings for source speaker selection and its potential impact on the VC model's accuracy.
\end{enumerate}

\section{Dataset}
\label{sec:dataset}

We have collected the NKI-OC-VC dataset for the VC task, comprising Dutch pathological speech from 16 oral cancer (OC) speakers (10 male) who had undergone surgery for tongue tumours and 5 healthy control speakers (2 male).

We collected data from the subjects at a maximum of three time points (stages): before the surgery (T1), after the surgery (T2), and six months after the surgery (T3). The recordings themselves took place in the patient room during scheduled speech therapy sessions.

The subjects were asked to read the standard Dutch text ``Jorinde en Joringel" \cite{son01_eurospeech} during the recording session. The total duration of all speech recordings is approximately 2.5 hours. One recording session (speaker/stage) is 5 minutes on average. Each speaker recites 92 sentences in total per stage. In some cases, patients felt the experiment was difficult, in that case, we prematurely stopped the experiment.

The speech was recorded with the Roland R-09HR field recorder at 44.1 kHz sampling frequency and 24-bit depth. This was later downsampled to 16 kHz and quantised to 16-bit depth. For the experiments, we have partitioned the 92 sentences into training (78 utterances), development (7 utterances), and test sets (7 utterances).

In order to obtain ratings for the overall severity of the speech, the speech of each speaker was rated by five Dutch SLPs on a 5-point Likert scale, where 5 meant healthy, and 1 meant severe. The ratings had a very high interrater correlation ($r=0.88$). The obtained ratings were averaged per speaker which we treat as the severity of the speaker for the rest of the paper. The available stages and the severity of each speaker are indicated in Table~\ref{tab:dataset}.

\begin{table}[htbp]
\centering
\caption{Speaker breakdown of the NKI-OC-VC dataset. The severity of the speaker according to 5 SLPs is noted in the parentheses. The patients marked with an asterisk (*) had to end the experiment prematurely, therefore we do not have all the 92 utterances from them.}
\resizebox{0.9\columnwidth}{!}{%
\begin{tabular}{@{}lcccccc@{}}
\toprule
    \textbf{ID} & \textbf{Control} & \textbf{T1} & \textbf{T2}         & \textbf{T3}      & \textbf{Gender} \\ \midrule
PEAM &                  & \checkmark \, (4.6)    & \checkmark* \, (4.4) & \checkmark \, (4.4)         & M            \\
PGAF &                  & \checkmark \, (4.6)    & \checkmark \, (4.6)            & \checkmark \, (4.6)         & F          \\
PHNF &                  & \checkmark \, (4.6)    & \checkmark \, (4.2)            & \checkmark \, (4.6)         & F          \\
PIIM &                  & \checkmark \, (3.8)    & \checkmark \, (3.8)            &                  & M            \\
PRVM &                  & \checkmark \, (5)      & \checkmark \, (4.4)            & \checkmark \, (4.8)         & M            \\
PJSM &                  & \checkmark \, (5)      & \checkmark \, (3)              & \checkmark \, (3)           & M            \\
\midrule
RGCM &                  &             &                     & \checkmark \, (1.2)         & M            \\
RBEM &                  &             &                     & \checkmark \, (2.2)         & M            \\
RCIM &                  &             &                     & \checkmark \, (1)           & M            \\
RIFF &                  &             &                     & \checkmark \, (1)           & F          \\
RKKM &                  &             &                     & \checkmark* \, (1.4)     & M            \\
RMKM &                  &             &                     & \checkmark \,(1)            & M            \\
RMRM &                  &             &                     & \checkmark \, (1.8)         & M            \\
ROEF &                  &             &                     & \checkmark \, (1.4)         & F          \\
RQNF &                  &             &                     & \checkmark* \, (1) & F          \\
RQOF &                  &             &                     & \checkmark \, (1.6)         & F          \\
\midrule
VAHM & \checkmark                &             &                     &                  & M            \\
VDSF & \checkmark                &             &                     &                  & F          \\
VMSM & \checkmark                &             &                     &                  & M            \\
VODF & \checkmark                &             &                     &                  & F          \\
VQBF & \checkmark                &             &                     &                  & F          \\
\bottomrule
\end{tabular}%
}
\label{tab:dataset}
\end{table}

\section{Proposed framework}
\label{sec:methods}

Given a speech sample from a source OC speaker (T2 or T3), and a speech sample from a target speaker before the surgery (T1), the VC aims to change the speaker identity of the speaker so that it resembles the T1 speaker and preserves the severity of the source T2 or T3 speech. In the following sections, we describe the Baseline and Proposed systems for the H2P-VC task.

\subsection{Baseline: PPG}

For the baseline, we follow the phonetic posteriorgram-based (PPG) recognition-synthesis approach implemented in S3PRL-VC\footnote{\url{https://github.com/unilight/s3prl-vc}}. 
A phonetic posteriorgram (PPG) is a feature map derived from an automatic speech recogniser that represents the posterior probabilites of phonetic units over time frames of the speech. We make adaptations so that the system is able to deal with Dutch speech \cite{huang2022s3prl}. 

In order to extract appropriate PPG features, we first train a Dutch Conformer E2E model ASR on the standard training set of the Corpus Gesproken Nederlands (CGN) dataset \cite{leeuwen2009results}. CGN contains Dutch recordings spoken by 1185 female and 1678 male speakers (age range 18-65 years old) from all over the Netherlands and Flanders. The Conformer models parameters were taken from \cite{guo2021recent,karita2019transformer_espnet}: 12 encoder layers and 6 decoder layers, all with 2048 dimensions; the attention dimension is 512 and the number of attention heads is 8; the convolution subsampling layer in the encoder has 2-layer CNNs with 256 channels, stride with 2, and a kernel size of 3. A default conv kernel size of 31 was used. Subword units with a vocabulary size of 5000 were used as basic units. After training, phonetic posteriorgram (PPG) features are extracted using the Conformer encoder.

As for incorporating speaker-specific information, we use the pre-trained ECAPA-TDNN x-vector checkpoint from the SpeechBrain toolbox with default parameters\footnote{\url{https://huggingface.co/speechbrain/spkrec-ecapa-voxceleb}} \cite{ravanelli2021speechbrain}. The Taco2-AR architecture used is described in \cite{huang2022s3prl}. Due to the scarcity of training data, the NKI-OC-VC dataset (around 2 hours), we train the model with a curated (RDH-VL) subset of the Mozilla Common Voice dataset\footnote{\url{https://github.com/r-dh/dutch-vl-tts}} \cite{ardilacommon} alongside the NKI-OC-VC dataset. The RDH-VL dataset contains 12.3 hours of speech from a Flemish Dutch speaker.

\subsection{Proposed: PPG + GST}

While PPGs are highly effective speaker-independent features, using PPGs and x-vectors only has several shortcomings with pathological speech. First, PPGs primarily capture time-variant features of pathological speech, and are unable to model time-invariant aspects. Second, if the ASR used to extract the PPGs can recognise pathological speech very well, the approach of PPG can lead to enhanced speech \cite{serrano2019parallel, zheng2022phonetic}. X-vectors can indeed model time-invariant aspects but x-vectors are trained only with healthy data, therefore their performance is likely insufficient.

In order to address this, we propose a global style-token (GST) based reference encoder to the architecture which we use as a severity encoder. GST has been proposed in \cite{wang2018style} as a way to control various aspects of prosody, such as speed, style of speaking, and style of singing \cite{valle2020mellotron, wu2019end}. It has been successfully used as a style transfer technique also in VC \cite{liu2020transferring}. We think that a mixture of using PPG to model linguistic information, x-vector to model severity information, and GST to model severity information is an efficient way to learn severity-related features from our dataset.

For the implementation of the GST reference encoder used as the severity encoder, we follow the default parameters of \cite{wang2018style} with the following exceptions: instead of adding the style embedding, we concatenate the style embedding, the x-vector and the encoder's hidden states, and add a linear projection layer. Furthermore, as the reference encoder leaks noise from the samples, we enhanced the speech with ConvTASNet using a public implementation\footnote{\url{https://github.com/asteroid-team/asteroid}} \cite{luo2019conv}.

HifiGAN was used as a neural vocoder for both systems \cite{kong2020hifi}. We followed an open-source implementation\footnote{\url{https://github.com/kan-bayashi/ParallelWaveGAN}}. The training data of the HifiGAN was the RDH-VL dataset.

\section{Experimental setup}

In the next sections, we present the methodology to evaluate three aspects of the systems: severity preservation, naturalness, and speaker identity. Due to a large number of possible conversion pairs, we only evaluate a limited set of converted speech. The selection of these sets is explained in their respective section. The audio samples can be found online\footnote{\url{https://h2pvc.github.io/}},

\subsection{Objective evaluation metrics}

\begin{table*}[t]
\centering
\caption{Results of P-ESTOI-based severity evaluation. Rows include the source speaker. Target speaker is always RDH-VL. \textbf{Boldface} indicates higher correlation, and better severity preservation.}
\resizebox{\textwidth}{!}{%
\begin{tabular}{lrrrrrrrrrrrrl}
\hline
                 & \multicolumn{1}{l}{$PGAF_{T2}$} & \multicolumn{1}{l}{$PHNF_{T2}$} & \multicolumn{1}{l}{$PIIM_{T2}$} & \multicolumn{1}{l}{$PJSM_{T2}$} & \multicolumn{1}{l}{$PRVM_{T2}$} & \multicolumn{1}{l}{$RBEM_{T3}$} & \multicolumn{1}{l}{$RCIM_{T3}$} & \multicolumn{1}{l}{$RGCM_{T3}$} & \multicolumn{1}{l}{$RMKM_{T3}$} & \multicolumn{1}{l}{$RMRM_{T3}$} & \multicolumn{1}{l}{$ROEF_{T3}$} & \multicolumn{1}{l}{$RQOF_{T3}$} & $r_{GT}$                               \\ \hline
$\text{P-ESTOI}_{PPG}$ & 1.00                        & 0.23                        & 0.14                        & 0.21                        & 0.37                        & 0.29                        & 0.08                        & 0.25                        & 0.13                        & 0.22                        & 0.16                        & 0.29                        & \multicolumn{1}{r}{0.49}          \\
$\text{P-ESTOI}_{PPG-GST}$ & 0.33                        & 0.30                        & 0.15                        & 0.27                        & 0.40                        & 0.23                        & 0.13                        & 0.22                        & 0.14                        & 0.31                        & 0.20                        & 0.30                        & \multicolumn{1}{r}{\textbf{0.90}} \\ \hline
$\text{P-ESTOI}_{GT}$  & 0.27                        & 0.32                        & 0.08                        & 0.18                        & 0.29                        & 0.23                        & 0.12                        & 0.17                        & 0.07                        & 0.27                        & 0.19                        & 0.25                        & -                                  \\ \hline
\end{tabular}%
}
\label{table:pestoi}
\end{table*}

\subsubsection{Severity: P-(E)STOI}
\label{subsec:stoi}

We use P-(E)STOI as an objective severity measure as it has been demonstrated to work well for the objective evaluation of dysarthric speech, and used previously to evaluate the severity of synthetic speech samples \cite{janbakhshi2019pathological}. We calculate the P-(E)STOI scores for each pathological ground truth (GT) utterance, using the corresponding utterance of the control speakers as a healthy reference. These scores are then averaged to provide a speaker-level score. We repeat this with a limited set of the VC utterance: we convert one post-operative (T2 if available) stage of each speaker to the target speaker in the RDH-VL dataset.  We compare the obtained P-(E)STOI scores for each VC system to the ground truth samples by calculating the Pearson's correlation ($r_{GT}$). 

Before doing our evaluation, we performed a preliminary analysis to ensure that P-ESTOI is indeed appropriate for severity evaluation, by calculating the P-(E)STOI scores on the GT samples first, and calculating the correlation between the scores and the severity labels. We found that both methods have a high correlation with the severity labels but the P-ESTOI had higher ($r = 0.82$) than the P-STOI ($r = 0.69$), therefore we will only consider P-ESTOI for the rest of our work.

\begin{table}[h]
\caption{Comparison of phoneme error rate (PER\%) of the two systems. \textbf{Boldface} indicates when the PER is closer to GT. \textit{Italics} indicate when the system overenhances the speech over the GT.}
\centering
\resizebox{0.7\columnwidth}{!}{%
\begin{tabular}{@{}lrrr@{}}
\toprule
        & \multicolumn{1}{l}{PPG} & \multicolumn{1}{l}{PPG-GST} & \multicolumn{1}{l}{GT} \\ \midrule
$PGAF_{T2}$    & \textit{47.86}          & \textbf{56.42}     & 52.92         \\
$PHNF_{T2}$    & \textbf{40.86}          & 45.53              & 38.13                  \\
$PIIM_{T2}$    & 84.44                   & \textbf{83.66}              & 80.54         \\
$PJSM_{T2}$    & 54.47                   & \textbf{44.36}              & 48.64         \\
$PRVM_{T2}$    & \textbf{41.25}          & 45.91             & 41.25                  \\
$RBEM_{T3}$    & \textit{67.70}          & \textbf{69.65}     & 74.32         \\
$RCIM_{T3}$    & \textit{92.61}          & \textbf{114.79}    & 105.45       \\
$RGCM_{T3}$    & \textit{80.16}          & \textbf{84.44}     & 82.88         \\
$RMKM_{T3}$    & \textit{84.44}          & \textit{\textbf{85.21}}     & 85.60         \\
$RMRM_{T3}$    & 61.48                   & \textbf{59.53}              & 58.37         \\
$ROEF_{T3}$    & \textit{77.04}          & \textit{\textbf{88.33}}     & 90.27         \\
$RQOF_{T3}$    & \textit{\textbf{68.87}} & \textit{68.48}     & 77.04        \\ \midrule
Average & \textit{66.76}                   & \textbf{70.53}              & 69.62                 
\end{tabular}%
}
\label{tab:per}
\end{table}

\subsubsection{Severity: Phoneme error rate (PER)}
\label{subsec:per}

We use the phoneme error rate (PER) as another measure for evaluating the severity objectively. We use a publicly available implementation\footnote{\url{https://huggingface.co/Clementapa/wav2vec2-base-960h-phoneme-reco-dutch}} phoneme recogniser which was trained on the Common Voice dataset \cite{ardilacommon}. We used \textit{phonemizer} to acquire phonetic transcriptions \cite{Bernard2021}. For the PER evaluation experiment, we convert the T2 (or T3 if T2 is not available) stage of each speaker to the target speaker in the RDH-VL dataset, and we calculate the PER per speaker and per system.

\subsection{Subjective evaluation experiments}

\subsubsection{Severity and speaker identity}
\label{subsec:sev_spk}
In evaluating speaker similarity, we adapt the protocol outlined in \cite{illa2021pathological} and incorporate an additional question about the similarity of speech severity. 
During the listening test, listeners are presented with two samples (either converted or ground truth), and are asked to judge (a) whether the two samples are from the
same speaker, (b) whether the two samples have the same severity of speech. Listeners are asked to rate the similarity between the two voices on a 4-point Likert scale, where 1 means 'not similar at all' and 4 means 'very similar'. For analysis, we converted these Likert scale ratings into percentages, i.e.,  a rating of '1' was considered as 0\%, and '4' as 100\%.
Due to the time constraints of the test, we limit our analysis to 5 speaker pairs here. The test was done by 3 Dutch SLPs.
For the listening experiment, we have prioritised pairs of source and target speakers who have similar severity. Choosing speakers with similar severity allows us to investigate whether the current practice of matching speakers based on their severity ratings is an appropriate approach for this VC task. Furthermore, compared to the protocol in \cite{huang2022towards}, this setup has the advantage that the speaker identity can also be evaluated with expert listeners.

\begin{table*}[]
\centering
\caption{Results of the subjective severity and identity evaluation experiments by 3 SLPs. S2S means similarity to source, S2T means similarity to target, and S-T means comparison of source to target. \textbf{Boldface} indicates better severity preservation.}
\resizebox{\textwidth}{!}{%
\begin{tabular}{lllllll|lllll}
\toprule
                 &                  & \multicolumn{4}{c}{\textbf{Severity}}                                                                 & \multicolumn{1}{c}{}    & \multicolumn{4}{c}{\textbf{Identity}}                                                                 & \multicolumn{1}{c}{}    \\ \midrule
                 &                  & \multicolumn{2}{c}{\textbf{PPG}}                  & \multicolumn{2}{c}{\textbf{PPG-GST}}              & \multicolumn{1}{c}{}    & \multicolumn{2}{c}{\textbf{PPG}}                  & \multicolumn{2}{c}{\textbf{PPG-GST}}              & \multicolumn{1}{c}{}    \\
\multicolumn{2}{c}{Conversion pair (Severity)} & \multicolumn{1}{c}{S2S} & \multicolumn{1}{c}{S2T} & \multicolumn{1}{c}{S2S} & \multicolumn{1}{c}{S2T} & \multicolumn{1}{c}{S-T} & \multicolumn{1}{c}{S2S} & \multicolumn{1}{c}{S2T} & \multicolumn{1}{c}{S2S} & \multicolumn{1}{c}{S2T} & \multicolumn{1}{c}{S-T} \\ \midrule
\multicolumn{2}{c}{$PIIM_{T2} \ (3.8) \rightarrow PJSM_{T2} \ (3)$}          & 57$\pm$19\%                  & 32$\pm$16\%                  & \textbf{84$\pm$12\%}                  & 11$\pm$11\%                  & 14$\pm$19\%                  & 3$\pm$6                    & 43$\pm$18\%                  & 81$\pm$14\%                  & 5$\pm$5\%                    & 0$\pm$0\%                    \\
\multicolumn{2}{c}{$PRVM_{T2} \ (4.4) \rightarrow  PHNF_{T2} \ (4.2)$}          & 43$\pm$18\%                  & 70$\pm$15\%                  & 43$\pm$18\%                  & 59$\pm$13\%                  & 28$\pm$19\%                  & 0$\pm$0\%                    & 48$\pm$16\%                  & 24$\pm$17\%                  & 48$\pm$15\%                  & 0$\pm$0\%                    \\
\multicolumn{2}{c}{$PHNF_{T2} \ (4.2) \rightarrow PRVM_{T2} \ (4.4)$}          & 49$\pm$18\%                  & 63$\pm$13\%                  & \textbf{59$\pm$19\%}                  & 49$\pm$17\%                  & 28$\pm$19\%                  & 0$\pm$0\%                    & 52$\pm$18\%                  & 38$\pm$19\%                  & 25$\pm$16\%                  & 0$\pm$0\%                    \\
\multicolumn{2}{c}{$PHNF_{T2} \ (4.2) \rightarrow PJSM_{T2} \ (3)$}          & 41$\pm$19\%                  & 57$\pm$15\%                  & \textbf{59$\pm$15\%}                  & 38$\pm$16\%                  & 44$\pm$27\%                  & 0$\pm$0\%                    & 63$\pm$16\%                  & 52$\pm$18\%                  & 10$\pm$8\%                   & 0$\pm$0\%                    \\
\multicolumn{2}{c}{$PGAF_{T2} \ (4.6) \rightarrow PHNF_{T2} \ (4.2)$}         & 41$\pm$17\%                  & 60$\pm$12\%                  & \textbf{78$\pm$12\%}                  & 68$\pm$13\%                  & 31$\pm$25\%                  & 37$\pm$17\%                  & 65$\pm$17\%                  & 84$\pm$14\%                  & 92$\pm$8\%                   & 75$\pm$21\%                  \\ \midrule
\multicolumn{2}{c}{Ideal}         & 100\%                  & S-T               & 100\%                  & S-T                  & 100\%                  & 0\%                 & 100\%                  & 0\%                  & 100\%                   & 0\%                  \\ \bottomrule
\end{tabular}%
}
\label{tab:severity_identity}
\end{table*}

\subsubsection{Naturalness}
\label{subsec:mos}
In order to evaluate the naturalness of the VC samples, we carried out a mean opinion score (MOS) test. We followed the setup in \cite{huang2022towards}, which is a variant of the standard MOS test but with increments of 0.5. This change is made in order to avoid lower precision of our naturalness test with high severity speech. 
For this test, we chose 7 source speakers from our dataset, each with differing levels of speech severity to capture a wide range. We created 7 conversion pairs in total using each source speaker, and targeting the same speaker from the RDH-VL dataset. This approach was taken due to evidence that severity and naturalness are often conflated in listener perceptions \cite{illa2021pathological, huang2022towards, halpern2022manipulation}. The test included 153 utterances in total, with 7 VC utterances per conversion pair and per system, alongside their corresponding 7 source ground truth utterances and 6 utterances from the target RDH-VL speaker. These utterances were rated by 7 native Dutch listeners. 

\subsection{Analysis of the impact of speech pathology on the x-vectors}

To assess the impact of speech pathology on x-vectors, we create three separate distributions of similarity scores by calculating the cosine similarity of x-vectors. First, we do same-speaker pre-operative comparisons of the utterances (T1). Second, we do same-speaker different-severity comparisons (T1+T2). Finally, we do different-speaker comparisons (non-target). By separating these distributions, it is possible to quantify the impact of severity on the x-vectors. To quantify this, we calculate the Equal Error Rate (EER) for each group, a common performance measure in speaker verification studies. In the context of our H2P-VC task, the EER quantifies the robustness of the x-vector extraction in the presence of speech pathology. Specifically, a high EER might indicate that the x-vectors are significantly influenced by the severity of the pathology.

\section{Results and discussion}
\label{sec:results}

\subsection{Objective evaluations}

\subsubsection{Severity: P-ESTOI}
Table~\ref{table:pestoi} shows the results of the P-ESTOI experiment. We find that the correlation with the ground truth P-ESTOI scores $r_{GT}$ are higher for the PPG-GST-based system, which indicates that the PPG-GST-based system has a better ability to capture the severity of the speech.

\subsubsection{Severity: PER}

The PER results in Table~\ref{tab:per} show that the proposed PPG-GST system has a mean PER that is closer to the GT than the PPG. We can observe that the PER is only closer in the case of speaker $PHNF_{T2}$ and $PRVM_{T2}$, and $RQOF_{T3}$. $PHNF_{T2}$ and $PRVM_{T2}$ are both speakers with relatively healthy ratings (4+), where we would expect severity-related modelling of the PPG-GST to be less beneficial. The difference in the case of $RQOF_{T3}$ is negligible. Furthermore, we observe that the PPG on average has a tendency to make the speech more intelligible than the ground truth, the average PER being nearly 3\% better than in the GT case. We conclude that the PPG-GST has a better severity preservation ability than the PPG based on the results of the PER experiment.

\subsection{Subjective evaluations}

\subsubsection{Severity and speaker identity}
The left side of Table~\ref{tab:severity_identity} shows the results of the severity experiment. The raters found that the severity of the GT source and target speaker sounded fairly different (S-T column), even though they were noted to have similar severity of speech. We conclude that using severity labels alone is insufficient for the source speaker selection based on this result. Therefore, for evaluating the severity preservation property, we should look at the S2S column rather than the S2T column. We find that for nearly all speakers the proposed model had a higher similarity to the source speakers' speech severity, with the exception of $PRVM_{T2} \rightarrow PHNF_{T2}$ where performance was on par.  We conclude the PPG-GST outperforms the baseline model in severity preservation. 

The right side of Table~\ref{tab:severity_identity} shows the results of the speaker identity conversion experiment.  We find that with the exception of the conversion pair $PGAF_{T2}\rightarrow PHNF_{T2}$, (S-T column) the raters rated the two speakers as different. Furthermore, the proposed system does not seem to achieve the same level of speaker identity conversion as the PPG system but outperforms the PPG in the case of the $PGAF_{T2} \rightarrow PHNF_{T2}$, and is on par in case of the $PRVM_{T2} \rightarrow PHNF_{T2}$ conversion pair. 

The challenges in evaluating speaker identity, as outlined in \cite{huang2022towards}, warrant further discussion. Even though we have acquired a parallel dataset of healthy and pathological speech from the same speaker, the precise severity of the speech pathology could not exactly be matched, despite our best efforts. As a result, it is still possible that differences in severity could influence the evaluation of speaker identity.

\subsubsection{Naturalness}

The results of the naturalness test are shown in Table~\ref{tab:mos}. Consistent with \cite{illa2021pathological,huang2022towards, halpern2022manipulation} , the high correlation between the severity labels and the GT naturalness ratings ($r=0.88$) shows that the listeners cannot differentiate well between these aspects.
Overall, the PPG-GST is rated as less natural than the PPG. However, this is not unexpected due to the following reasons: (1) The better preservation of severity is perceived as lower naturalness. (2) The reference encoder in the PPG-GST directly interacts with the noisy filterbank during reconstruction, leading to noisier sounding examples.

\begin{table}[]
\centering
\caption{Mean opinion scores (MOS) from the naturalness experiment from 7 native listeners.}
\resizebox{0.8\columnwidth}{!}{%
\begin{tabular}{@{}lcccc@{}}
\toprule
        & \multicolumn{1}{l}{PPG} & \multicolumn{1}{l}{PPG-GST} & \multicolumn{1}{l}{GT} & \multicolumn{1}{l}{Severity} \\ \midrule
$PGAF_{T1}$ & 3.66                    & 3.31                        & 4.65                   & 4.6                          \\
$PHNF_{T2}$ & 3.58                    & 2.72                        & 4.47                   & 4.6                          \\
$PIIM_{T1}$ & 2.89                    & 2.63                        & 3.79                   & 3.8                          \\
$RBEM_{T3}$ & 2.55                    & 2.42                        & 3.19                   & 2.2                          \\
$RMRM_{T3}$ & 2.37                    & 2.16                        & 3.90                   & 1.8                          \\
$RQOF_{T3}$ & 2.51                    & 2.18                        & 3.13                   & 1.6                          \\
$RCIM_{T3}$ & 2.09                    & 1.86                        & 2.99                   & 1                            \\
RDH-VL  & \multicolumn{1}{l}{}    & \multicolumn{1}{l}{}        & 4.54                   & \multicolumn{1}{l}{}         \\ \midrule
Average & 2.81                    & 2.47                        & 3.73                   & \multicolumn{1}{l}{}         \\ \bottomrule
\end{tabular}%
}
\label{tab:mos}
\end{table}

\subsection{Analysis of the impact of speech pathology on x-vectors}
\begin{figure}
    \includegraphics[width=\columnwidth]{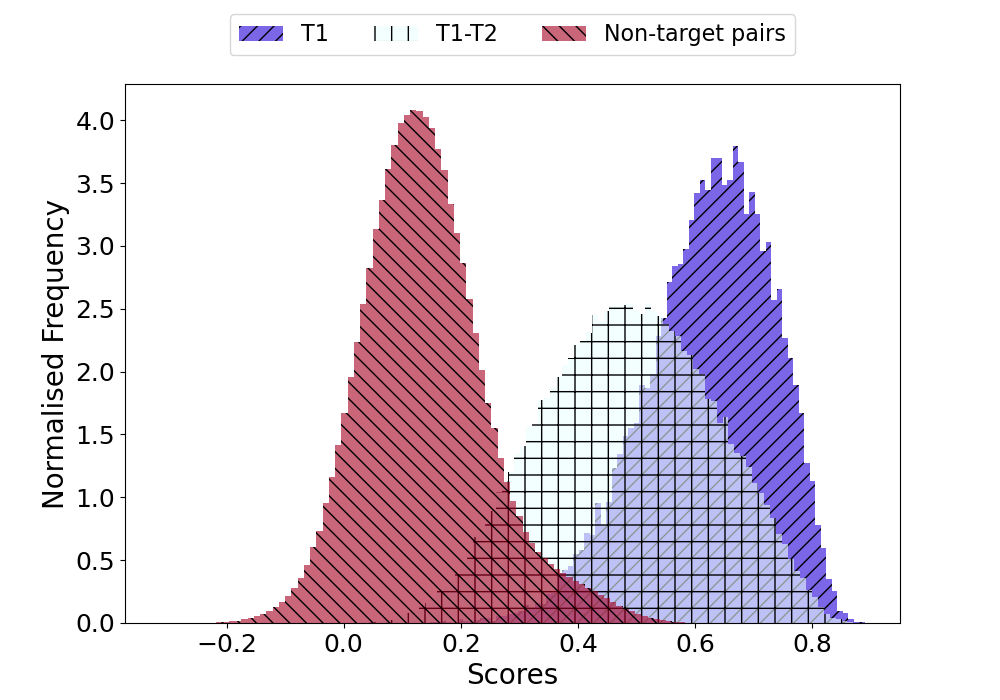}
    \caption{Impact of pathology on the extracted x-vectors. Best viewed in colour.}
    \label{fig:speaker_similarity}
\end{figure}

Figure~\ref{fig:speaker_similarity} shows the impact of speech pathology on the similarity of x-vectors. As expected, when looking only at the T1 speakers (T1), there is a high similarity as we are comparing the same speakers with the same identity. Introducing the T2 speakers into the comparison, the similarity decreases (T1+T2) but it is still much higher than the non-target scores. This shows that pathology impacts embeddings.

Table~\ref{tab:eer_embedding} quantifies the impact with EERs, and also shows the impact of speech pathology on the speaker-level. We observe that adding the T2 utterances always makes the EER higher. Including all speakers in the analysis, an EER of 8.87\% can be maintained with even the T2 distribution. When looking at the results on the speaker level, the worst EER (16.57\%) is achieved for PIIM, and the best EER is achieved for PHNF (3.70\%). Interestingly, there does not seem to be a clear pattern in the increase in severity, and the increase in EER. The lack of this relation seems to be consistent with the finding of \cite{arasteh2022effect}, which reports that the effect of speech pathology lies in the range of other factors in speaker verification, such as the quality of the microphone. 

We can conclude that the x-vectors are impacted by the speech pathology, however, the speaker still remains identifiable. Therefore, the pre-operative (T1) x-vector should contain sufficient information about the identity.
We acknowledge that the current analysis has some limitations. First, we only have parallel data from low-to-mid severity speakers, therefore it might be that the extracted embeddings are affected for high severity speakers. Furthermore, as we use utterances from the same recording to calculate the T1 EER, the estimation of T1 identification performance is likely somewhat optimistic.

\subsection{Summary of Results and Comparative Analysis of PPG and PPG+GST}

In the previous sections, we have provided extensive evaluation of the PPG and PPG+GST systems. The baseline PPG system demonstrates better performance in terms of speaker identity conversion, as noted in the subjective identity evaluation. However, this assessment could be influenced by the severity of speech pathology. Furthermore, the PPG system falls short of preserving the severity of the speech, especially when compared to the PPG+GST system. Finally, as we have seen by the PER results, it tends to overenhance the speech, potentially leading to a less realistic output.

On the other hand, the proposed PPG+GST system preserves the severity of speech well during conversion as shown by the objective P-ESTOI, PER and subjective evaluations. In terms of speaker identity conversion, however, the PPG+GST system performs slightly less well than the PPG system. In addition, the PPG+GST system may be prone to introducing noise in the output, which can detract from its naturalness. In summary, we find that while PPG+GST still needs further work, however, its superior ability to preserve speech severity makes PPG-GST a more appropriate choice for mid-to-high severity voices.

\begin{table}[]
\centering
\caption{Equal error rates (\%) of T1 and T1-T2 distributions.}
\resizebox{\columnwidth}{!}{%
\begin{tabular}{@{}lrrrrrrr@{}}
\toprule
   & \multicolumn{1}{l}{PEAM} & \multicolumn{1}{l}{PGAF} & \multicolumn{1}{l}{PHNF} & \multicolumn{1}{l}{PRVM} & \multicolumn{1}{l}{PIIM} & \multicolumn{1}{l}{PJSM} & \multicolumn{1}{l}{All} \\ \midrule
T1 & 3.06\%                   & 1.17\%                   & 1.15\%                   & 1.89\%                   & 2.96\%                   & 3.92\%                   & 2.59\%                  \\
T1-T2 & 10.05\%                  & 7.46\%                   & 3.70\%                   & 10.25\%                  & 16.57\%                  & 7.76\%                   & 8.87\%                  \\ \bottomrule
\end{tabular}%
}
\label{tab:eer_embedding}
\end{table}

\begin{figure*}
    \centering
    \includegraphics[width=\textwidth]{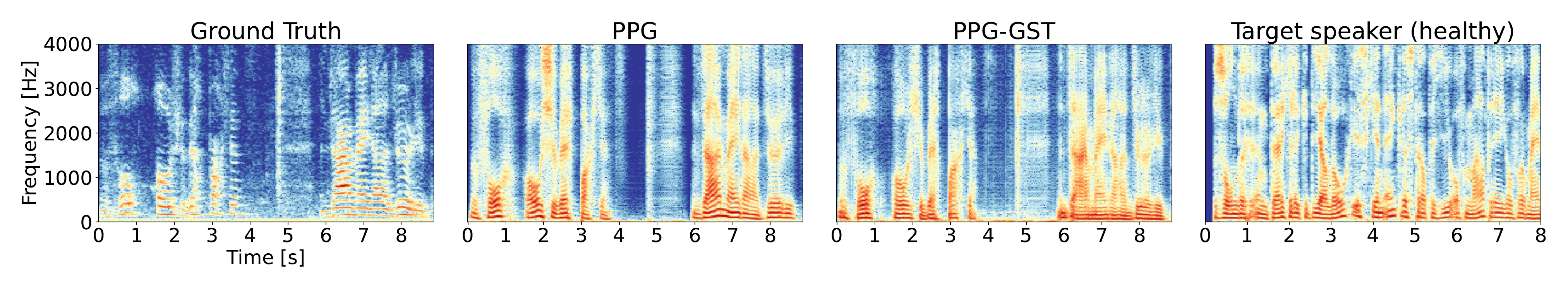}
    \vspace{-3em}
    \caption{Spectrogram comparison (left to right) ground truth, baseline, proposed and healthy speaker identity (target). Best viewed in colour.}
    \label{fig:samples}
\end{figure*}

\section{Conclusion}
\label{sec:conclusion}

In this paper, we presented a new voice conversion system which utilises phonetic posteriorgram features and global style tokens to preserve the speech pathology while converting the speaker identity. 

\textbf{Our system is able to preserve the voice severity better than the baseline while still resembling the target speaker according to our subjective and objective evaluations}. The naturalness of our system is lower than the baseline, however, this is most likely because more severe speech is perceived as less natural. Nevertheless, there is still a large gap between the naturalness of ground truth and converted pathological speech. We think that techniques to better disentangle speech severity and noise should be one focus of future research.

\textbf{Speaker identity is impacted by severity but usually not all speaker information is lost.} This means that it is feasible to use the pre-operative x-vector to convert to post-operative speaker's characteristics. In future work, we are interested if pathology-robust extraction of x-vectors could result in better conversion quality. One possible approach for this would be to train an ECAPA-TDNN with parallel pathological and healthy speech samples. 

\textbf{Our findings also show that choosing source speakers based on severity labels alone is insufficient}. The variability of pathological speech makes it challenging to use a broad severity label for accurate speaker selection. This observation emphasises the need for more nuanced metadata (e.g. age, gender, other therapeutic variables) collection from speakers.

\section{Acknowledgements}
\label{sec:ack}
The data collection in the paper received ethical approval under the number IRBd20-159. We would like to thank the speech language pathologists (Klaske van Sluis, Lisette van der Molen, Marise Neijman, Rianne van Echtelt-Willig, Merel Latenstein) who agreed to participate in the evaluation experiments. The Department of Head and Neck Oncology and surgery of the Netherlands Cancer Institute receives a research grant from Atos Medical (H\"orby, Sweden), which contributes to the existing infrastructure for quality of life research. This work was supported in part by JST CREST Grant Number JPMJCR19A3, Japan.

\bibliographystyle{IEEEbib}
\bibliography{mybib}

\end{document}